\documentclass[floatfix,twocolumn,showpacs,prd,aps,tightenlines,superscriptaddress]{revtex4-1}
\usepackage{graphicx}
\usepackage{enumerate}
\usepackage{color}
\usepackage{psfrag}
\usepackage{dcolumn}
\usepackage{bm}

\usepackage{amssymb}
\usepackage{amsmath}
\usepackage{amsfonts}
\usepackage{longtable}
\usepackage{xspace}
\usepackage{afterpage} 
\usepackage{mathrsfs}

\usepackage{array}
\newcolumntype{H}{>{\lrbox0}c<{\endlrbox}@{}}

\def\eatcell#1\unskip{}
\newcolumntype{E}{>{\eatcell}c@{}}

\usepackage{collcell}
\makeatletter
\newcolumntype{G}{>{\collectcell\@gobble}c<{\endcollectcell}@{}}
\makeatother

\newcommand{\beq}{\begin{equation}}
\newcommand{\eeq}{\end{equation}}

\newcommand{\be}{\begin{equation}}
\newcommand{\ee}{\end{equation}}
\newcommand{\bea}{\begin{eqnarray}}
\newcommand{\eea}{\end{eqnarray}}
\newcommand{\bes}{\begin{subequations}}
\newcommand{\ees}{\end{subequations}}

\usepackage{bm}

\begin{document}

\title{Quasicircular Orbital Parameters for Numerical Relativity Revisited}

\author{Alessandro Ciarfella}
  \author{James Healy}
  \author{Carlos O. Lousto} 
\affiliation{Center for Computational Relativity and Gravitation,
School of Mathematical Sciences,
Rochester Institute of Technology, 85 Lomb Memorial Drive, Rochester,
New York 14623, USA}
  \author{Hiroyuki Nakano}
\affiliation{Faculty of Law, Ryukoku University, Kyoto 612-8577, Japan}

\date{\today}

\begin{abstract}
In the post-Newtonian (PN) expansion, 
we extend the determination of quasicircular
orbital parameters to be used by subsequent full numerical simulations
to the 3.5PN order, and find that this
leads to lower eccentricities, $e$, than with our previous method 
that used up to 3PN order.
We also supplement the computation of the radial infall due to radiation reaction
and the location of the center of mass to 3.5PN order, providing explicit formulas.
In addition, we consider the small mass ratio limit by explicitly including the
Schwarzschild and Kerr limits, the later in quasi-isotropic as well as in our
standard use of ADMTT coordinates. We evolve binaries with a $q=1/16$ 
mass ratio by using 3PN, 3.5PN, 3.5PN+Schwarzschild, 3.5PN+KerrQISO and 3.5PN+KerrADMTT quasicircular data for three different configurations
where the larger hole intrinsic spins are $\chi^z=-0.8$, $-0.4$ and $+0.8$.
Using different measures
of eccentricity from the black hole trajectories and from the waveform amplitudes
and phases, we determine a systematic reduction of eccentricities with respect to
the 3PN initial values
by factors of up to an order of magnitude, and reaching the desired
$e\sim10^{-3}$ threshold.
\end{abstract}

\maketitle

\section{Introduction}\label{sec:Intro}

The breakthroughs in numerical relativity (NR) \cite{Pretorius:2005gq,Campanelli:2005dd,Baker:2005vv}
have enabled the detailed modeling of gravitational waves from the merger of binary black holes (BBHs), predictions eventually first detected ten years afterwards by LIGO \cite{LIGOScientific:2016aoc}.
Since then, improvements in the modeling of gravitational waves \cite{LISAConsortiumWaveformWorkingGroup:2023arg}
and the creation of full numerical simulations catalogs \cite{Boyle:2019kee,Healy:2022wdn,Ferguson:2023vta}
have gone along with the detection of over a hundred BBH mergers
\cite{KAGRA:2021vkt}.

To represent the most likely configurations
at the late inspiral stages of the binaries, quasicircular (QC) initial parameters
have been chosen to create the first round of simulations to populate
NR catalogs. Historically,
two main approaches have been used in order to achieve the initial
parameters leading to QC orbits. The first was the
minimization of an effective potential \cite{Cook:1994va} which
lead to families of close binaries
successfully evolved in the Lazarus approach \cite{Baker:2002qf},
previous to the 2005 breakthroughs. It was then realized that the
post-Newtonian (PN) approach lead to an alternative approximation to
QC orbits \cite{Baker:2002qf}. Those studies were extended
and formalized to third post-Newtonian (3PN) order in Ref.~\cite{Healy:2017zqj}
and served to produce low eccentricity simulations
that populated the RIT catalogs \cite{Healy:2017psd,Healy:2019jyf,Healy:2020vre,Healy:2022wdn} for comparable mass binaries, then used directly to obtain independently
parameter estimations for the O1/O2 LIGO observations \cite{Healy:2020jjs}.

In this paper, we extend further this approach for the determination of
QC parameters to the 3.5PN order to provide even lower eccentricities,
$e\leq10^{-3}$, suitable for modeling sources for the higher sensitivity third
generation (3G) gravitational wave detectors and for the intermediate to small
mass ratio binaries in the
sensitivity range of the LISA space detector. In fact, as we move from
comparable mass ratios ($1/10<q\leq1$) to intermediate ones
($1/100<q\leq1/10$) \cite{Lousto:2022hoq},
the computational requirements to
perform simulations from a given initial orbital frequency (or
binary's separation) increases as $\sim q^{-2}$, thus making it steeply
costly. In practice, we will perform simulations from closer distances
(thus decreasing computational costs to $\sim q^{-1}$) making it necessary
to include higher PN orders in the QC orbit
determinations, and possibly including the particle limit behavior in
the Schwarzschild and Kerr backgrounds when $q \ll 1$.

These improved initial low eccentricity determinations also benefit
the iterative approaches \cite{Pfeiffer:2007yz,Buonanno:2010yk}
since they require evolutions of a couple of orbits for each iteration
(becoming costly in the $q \ll 1$ regime), hence a lower initial $e$ seed
reduces the number of needed iterations to achieve a given level of $e\approx0$.

\section{Approximate initial quasicircular orbits}\label{sec:ID}

Here, we follow the procedure developed in Ref.~\cite{Healy:2017zqj}
to obtain QC orbits, extending it to the full 3.5PN order,
including all spin terms, in the Arnowitt-Deser-Misner
transverse traceless (ADMTT) gauge that we found to reproduce
well the initial coordinates chosen in our full numerical simulations.
We will later proceed to supplement this approach by including
the particle limit in the Schwarzschild and Kerr background cases,
with applications to intermediate to small mass ratio BBH simulations.

\subsection{3.5PN Hamiltonian}\label{3.5PN Hamiltonian}

From Ref.~\cite{Schafer:2018kuf} the Hamiltonian at 3.5PN order in the
 ADMTT gauge can be written as
\bea
H &=& H_{\rm N} + H_{\rm 1PN}+ H_{\rm 2PN}+ H_{\rm 3PN}
\cr &&
+ H_{\rm LOSO}+ H_{\rm NLOSO}+ H_{\rm NNLOSO}
\cr &&
+ H_{{\rm LO}S_1S_2}+H_{{\rm NLO}S_1S_2}
\cr &&
+ H_{{\rm LO}S^2}+ H_{{\rm NLO}S^2}+ H_{{\rm LO}S^3} ,
\label{Hamiltonian}
\eea
where each Hamiltonian in the right-hand side of the above equation
denotes Newtonian (N), 1PN, 2PN, 3PN, 
leading-order (LO) spin-orbit (SO), next-to-leading-order (NLO) SO,
next-to-next-to-leading-order (NNLO) SO,
LO $S_1$-$S_2$, NLO $S_1$-$S_2$, 
LO spin-squared ($S^2$), NLO $S^2$, and LO cubic-in-spin ($S^3$)
ones, respectively.

For the sake of simplicity,
we will assume that the
linear momenta of the black holes are opposed and
directed along the $y$-axis and that the holes lie along the $x$-axis in our initial configuration,
so that we can straightforwardly identify $P_\phi$ with $L_z$.

In absence of radiation reaction,
the QC conditions in polar coordinates read
\begin{equation}
P_r = 0 , \qquad \frac{\partial H}{\partial r} = 0.
\end{equation}
These conditions give us an equation for $P_{\phi} = L_z$.

We then evaluate the orbital frequency, $\Omega$, as 
\begin{equation}
\Omega = \frac{\partial H}{\partial P_\phi},
\end{equation}
the tangential momentum, $P_t$, as 
\begin{equation}
P_t = \frac{P_{\phi}}{r},
\end{equation}
and finally, the total ADM mass,
\begin{equation}
M_{\rm ADM} = M + H ,
\end{equation}
where $M=m_1+m_2$.

In the following, we will adopt the units $G=1=c$, and notation,
$q = m_2/m_1\leq1$ for the mass ratio, with $\eta = q/(1+q)^2$,
the intrinsic spins $\vec{\chi}_i = \vec{S}_i/m_i^2$, and $r$ is
the orbital distance between the two black holes.

The explicit results of evaluating the equations above, computed at 3.5PN order,
in terms of $\Omega$, are
\begin{widetext}
\bea
P_t &=& \frac{M q}{(1+q^2)}\Big\{(M \Omega)^{1/3}+\frac{(15+29 q+15 q^2) (M \Omega)}{6 (1+q)^2}-\frac{2 ((4+3 q) \chi_{1z}+q (3+4 q) \chi_{2z}) (M \Omega)^{4/3}}{3 (1+q)^2}
\cr &&
+\frac{1}{72 (1+q)^4}\big[9 (49-8
   \chi_{1x}^2+4 \chi_{1y}^2+4 \chi_{1z}^2)
\cr &&
-72 q (-20+2 \chi_{1x}^2-\chi_{1y}^2-\chi_{1z}^2+2 \chi_{1x} \chi_{2x}-\chi_{1y} \chi_{2y}-\chi_{1z} \chi_{2z})
\cr &&
   -9 q^4 (-49+8
   \chi_{2x}^2-4 \chi_{2y}^2-4 \chi_{2z}^2)
\cr &&
-72 q^3 (-20+2 \chi_{1x} \chi_{2x}+2 \chi_{2x}^2-\chi_{1y} \chi_{2y}-\chi_{2y}^2-\chi_{1z} \chi_{2z}-\chi_{2z}^2)
\cr &&
   +q^2 (1997-72
   \chi_{1x}^2+36 \chi_{1y}^2+36 \chi_{1z}^2-288 \chi_{1x} \chi_{2x}-72 \chi_{2x}^2+144 \chi_{1y} \chi_{2y}+36 \chi_{2y}^2
\cr && +144 \chi_{1z} \chi_{2z}+36 \chi_{2z}^2)\big] (M\Omega)^{5/3}
\cr &&
   -\frac{((16+29 q+22 q^2+7 q^3) \chi_{1z}+q (7+22 q+29 q^2+16 q^3) \chi_{2z}) (M \Omega)^2}{2 (1+q)^4}
\cr &&
   +\frac{1}{5184 (1+q)^6}\big[-36 (-2223+696
   \chi_{1x}^2-60 \chi_{1y}^2+4 \chi_{1z}^2)
\cr &&
   +36 q^4 (7480+483 \pi ^2+318 \chi_{1x}^2-372 \chi_{1y}^2-324 \chi_{1z}^2+636 \chi_{1x} \chi_{2x}-330 \chi_{2x}^2-2664 \chi_{1y}
   \chi_{2y}
\cr &&
-1296 \chi_{2y}^2-872 \chi_{1z} \chi_{2z}-1504 \chi_{2z}^2)
   -36 q^6 (-2223+696 \chi_{2x}^2-60 \chi_{2y}^2+4 \chi_{2z}^2)
\cr && +9 q (483 \pi ^2-4 (-6721+1368
   \chi_{1x}^2+372 \chi_{1y}^2+596 \chi_{1z}^2+576 \chi_{1y} \chi_{2y}+168 \chi_{1z} \chi_{2z}))
\cr &&
   +2 q^3 (13041 \pi ^2+2 (53681+5940 \chi_{1x}^2-11124 \chi_{1y}^2-11124
   \chi_{1z}^2+11448 \chi_{1x} \chi_{2x}
\cr &&
+5940 \chi_{2x}^2-37584 \chi_{1y} \chi_{2y}-11124 \chi_{2y}^2
   -12672 \chi_{1z} \chi_{2z}-11124 \chi_{2z}^2))
\cr &&
+36 q^2 (483 \pi ^2-2
   (-3740+165 \chi_{1x}^2+648 \chi_{1y}^2+752 \chi_{1z}^2
\cr &&
-318 \chi_{1x} \chi_{2x}-159 \chi_{2x}^2+1332 \chi_{1y} \chi_{2y}
   +186 \chi_{2y}^2+436 \chi_{1z} \chi_{2z}+162
   \chi_{2z}^2))
\cr &&+9 q^5 (483 \pi ^2-4 (-6721+1368 \chi_{2x}^2+576 \chi_{1y} \chi_{2y}+372 \chi_{2y}^2+168 \chi_{1z} \chi_{2z}
   +596 \chi_{2z}^2))\big] (M\Omega)^{7/3}
\cr &&+\frac{1}{432 (1+q)^6}\big[288 (1+q)^2 (2+3 q) \chi_{1z}^3+288 q (1+q)^2 (7+8 q) \chi_{1z}^2 \chi_{2z}
\cr &&
+\chi_{1z} (288 (-37+2 \chi_{1x}^2+2 \chi_{1y}^2)-72 q (355+2
   \chi_{1x}^2-28 \chi_{1y}^2+26 \chi_{1x} \chi_{2x}-40 \chi_{1y} \chi_{2y})
\cr &&
-3 q^3 (4951+432 \chi_{1x}^2-288 \chi_{1y}^2+2640 \chi_{1x} \chi_{2x}+6528 \chi_{2x}^2-3264 \chi_{1y}
   \chi_{2y}
   -3696 \chi_{2y}^2-5232 \chi_{2z}^2)
\cr &&
-18 q^4 (475+168 \chi_{1x} \chi_{2x}+1048 \chi_{2x}^2-192 \chi_{1y} \chi_{2y}-560 \chi_{2y}^2-928 \chi_{2z}^2)
\cr &&
   -72 q^5 (39+84\chi_{2x}^2-42 \chi_{2y}^2-82 \chi_{2z}^2)+q^2 (-24235-2016 \chi_{1x}^2+2304 \chi_{1y}^2
\cr &&-6768 \chi_{1x} \chi_{2x}-6768 \chi_{2x}^2+9216 \chi_{1y} \chi_{2y}
   +4032 \chi_{2y}^2+4896
   \chi_{2z}^2))
\cr &&+q \chi_{2z} (216 (-13+8 \chi_{1x}^2-4 \chi_{1y}^2)+18 q (-475+248 \chi_{1x}^2-160 \chi_{1y}^2-312 \chi_{1x} \chi_{2x}+48 \chi_{1y}
   \chi_{2y})
\cr &&
   -72 q^4 (355+134 \chi_{1x} \chi_{2x}+218 \chi_{2x}^2-40 \chi_{1y} \chi_{2y}-82 \chi_{2y}^2-82 \chi_{2z}^2)
\cr &&
-288 q^5 (37+22 \chi_{2x}^2-8 \chi_{2y}^2-8
   \chi_{2z}^2)
\cr &&
   +3 q^2 (-4951+1248 \chi_{1x}^2-1056 \chi_{1y}^2-6960 \chi_{1x} \chi_{2x}-1008 \chi_{2x}^2+1536 \chi_{1y} \chi_{2y}+432 \chi_{2y}^2+432 \chi_{2z}^2)
\cr &&
   +q^3
   (-24235+1008 \chi_{1x}^2-1152 \chi_{1y}^2-24912 \chi_{1x} \chi_{2x}-12384 \chi_{2x}^2
\cr &&
+6624 \chi_{1y} \chi_{2y}+4896 \chi_{2y}^2+4896 \chi_{2z}^2)))\big]
(M\Omega)^{8/3}\Big\},
   \eea
for the QC tangential linear momentum $P_t$,
\bea
M_{\rm ADM}&=&M + \frac{M q}{1+q^2}\Big\{-\frac{(M \Omega)^{2/3}}{2}+\frac{(9+19 q+9 q^2) (M \Omega)^{4/3}}{24 (1+q)^2}-\frac{((4+3 q) \chi_{1z}+q (3+4 q) \chi_{2z}) (M \Omega)^{5/3}}{3 (1+q)^2}
\cr &&
+\frac{1}{48 (1+q)^4}(81-48
   \chi_{1x}^2+24 \chi_{1y}^2+24 \chi_{1z}^2+q (267-96 \chi_{1x}^2+48 \chi_{1y}^2+48 \chi_{1z}^2-96 \chi_{1x} \chi_{2x}
\cr &&
+48 \chi_{1y} \chi_{2y}+48 \chi_{1z} \chi_{2z})
   +q^4 (81-48
   \chi_{2x}^2+24 \chi_{2y}^2+24 \chi_{2z}^2)
\cr &&
+q^2 (373-48 \chi_{1x}^2+24 \chi_{1y}^2+24 \chi_{1z}^2-192 \chi_{1x} \chi_{2x}-48 \chi_{2x}^2+96 \chi_{1y} \chi_{2y}+24 \chi_{2y}^2\cr &&
   +96\chi_{1z} \chi_{2z}+24 \chi_{2z}^2)+q^3 (267-96 \chi_{1x} \chi_{2x}-96 \chi_{2x}^2+48 \chi_{1y} \chi_{2y}+48 \chi_{2y}^2+48 \chi_{1z} \chi_{2z}+48 \chi_{2z}^2))
   (M \Omega)^2\cr &&
   -\frac{((72+140 q+96 q^2+27 q^3) \chi_{1z}+q (27+96 q+140 q^2+72 q^3) \chi_{2z}) (M \Omega)^{7/3}}{18 (1+q)^4}
\cr &&
   -\frac{1}{10368 (1+q)^6}\big[5 (9
   (-1215+960 \chi_{1x}^2+96 \chi_{1y}^2-416 \chi_{1z}^2)
\cr &&
-9 q (401+246 \pi ^2-2112 \chi_{1x}^2-1056 \chi_{1y}^2+736 \chi_{1z}^2
\cr &&
   -1152 \chi_{1y} \chi_{2y}+192 \chi_{1z}\chi_{2z})+9 q^6 (-1215+960 \chi_{2x}^2+96 \chi_{2y}^2-416 \chi_{2z}^2)
\cr &&
-9 q^2 (-9145+984 \pi ^2-816 \chi_{1x}^2-2496 \chi_{1y}^2
\cr &&
   -64 \chi_{1z}^2+288 \chi_{1x} \chi_{2x}+528
   \chi_{2x}^2-4800 \chi_{1y} \chi_{2y}-672 \chi_{2y}^2+832 \chi_{1z} \chi_{2z}-288 \chi_{2z}^2)\cr &&
   -9 q^4 (-9145+984 \pi ^2+528 \chi_{1x}^2-672 \chi_{1y}^2-288 \chi_{1z}^2+288 \chi_{1x}
   \chi_{2x}
\cr &&
-816 \chi_{2x}^2-4800 \chi_{1y} \chi_{2y}-2496 \chi_{2y}^2+832 \chi_{1z} \chi_{2z}\cr &&
   -64 \chi_{2z}^2)-9 q^5 (401+246 \pi ^2-2112 \chi_{2x}^2-1152 \chi_{1y} \chi_{2y}-1056
   \chi_{2y}^2+192 \chi_{1z} \chi_{2z}+736 \chi_{2z}^2)
\cr &&
   +q^3 (149951-13284 \pi ^2-7776 \chi_{1x}^2+19872 \chi_{1y}^2+6048 \chi_{1z}^2-5184 \chi_{1x} \chi_{2x}
\cr &&
-7776 \chi_{2x}^2+65664
   \chi_{1y} \chi_{2y}+19872 \chi_{2y}^2
   -11520 \chi_{1z} \chi_{2z}+6048 \chi_{2z}^2))] (M \Omega)^{8/3}
\cr &&
   +\frac{1}{24 (1+q)^6}\big[24 q (1+q)^2 \chi_{1z}^3+24 q (1+q)^2 (1+2 q)
   \chi_{1z}^2 \chi_{2z}+q \chi_{2z} (9 (-9+16 \chi_{1x}^2-8 \chi_{1y}^2)
\cr &&
   +12 q (-15+34 \chi_{1x}^2-20 \chi_{1y}^2-18 \chi_{1x} \chi_{2x})-24 q^4 (31+17 \chi_{1x}
   \chi_{2x}+29 \chi_{2x}^2-4 \chi_{1y} \chi_{2y}-10 \chi_{2y}^2-10 \chi_{2z}^2)\cr &&
   -12 q^5 (27+24 \chi_{2x}^2-8 \chi_{2y}^2-8 \chi_{2z}^2)
\cr && +12 q^2 (-19+32 \chi_{1x}^2-22
   \chi_{1y}^2-70 \chi_{1x} \chi_{2x}-10 \chi_{2x}^2+8 \chi_{1y} \chi_{2y}+4 \chi_{2y}^2+4 \chi_{2z}^2)
\cr &&
   +4 q^3 (-137+30 \chi_{1x}^2-24 \chi_{1y}^2-258 \chi_{1x} \chi_{2x}-132
   \chi_{2x}^2+48 \chi_{1y} \chi_{2y}+48 \chi_{2y}^2+48 \chi_{2z}^2))
\cr &&
   +\chi_{1z} (-324+96 \chi_{1x}^2+24 q (-31+7 \chi_{1x}^2+\chi_{1y}^2+\chi_{1x} \chi_{2x}+4 \chi_{1y}
   \chi_{2y})-3 q^5 (27+96 \chi_{2x}^2-48 \chi_{2y}^2-80 \chi_{2z}^2)
\cr &&
   -12 q^4 (15+6 \chi_{1x} \chi_{2x}+74 \chi_{2x}^2-12 \chi_{1y}\chi_{2y}-40\chi_{2y}^2-56\chi_{2z}^2)
\cr &&
-12 q^3 (19+2 \chi_{1x}^2-2 \chi_{1y}^2+10 \chi_{1x} \chi_{2x}+76 \chi_{2x}^2
-32 \chi_{1y} \chi_{2y}
   -44 \chi_{2y}^2-52 \chi_{2z}^2)
\cr &&
+4 q^2 (-137+12
   \chi_{1x}^2+12 \chi_{1y}^2-6 \chi_{1x} \chi_{2x}-78 \chi_{2x}^2
\cr &&+84 \chi_{1y} \chi_{2y}+48 \chi_{2y}^2+48 \chi_{2z}^2))\big] (M \Omega)^3\Big\},
\eea
for the total ADM mass of the system, and 
\bea\label{eq:r}
\frac{r}{M}&=& \frac{1}{(M \Omega)^{2/3}}-\frac{3+5 q+3 q^2}{3 (1+q)^2}-\frac{((4+3 q) \chi_{1z}+q (3+4 q) \chi_{2z}) (M \Omega)^{1/3}}{6 (1+q)^2}
\cr &&
+\frac{(M\Omega)^{2/3}}{72 (1+q)^4}\big[18 (-1-4 \chi_{1x}^2+2
   \chi_{1y}^2+2 \chi_{1z}^2)
\cr &&
-9 q (-1+16 \chi_{1x}^2-8 \chi_{1y}^2-8 \chi_{1z}^2+16 \chi_{1x} \chi_{2x}-8 \chi_{1y} \chi_{2y}-8 \chi_{1z} \chi_{2z})
\cr &&
   -9 q^3 (-1+16 \chi_{1x}
   \chi_{2x}+16 \chi_{2x}^2-8 \chi_{1y} \chi_{2y}-8 \chi_{2y}^2-8 \chi_{1z} \chi_{2z}-8 \chi_{2z}^2)-18 q^4 (1+4 \chi_{2x}^2-2 \chi_{2y}^2-2 \chi_{2z}^2)
\cr &&
   +q^2 (62-72
   \chi_{1x}^2+36 \chi_{1y}^2+36 \chi_{1z}^2-288 \chi_{1x} \chi_{2x}-72 \chi_{2x}^2+144 \chi_{1y} \chi_{2y}+36 \chi_{2y}^2+144 \chi_{1z} \chi_{2z}+36 \chi_{2z}^2)\big]
\cr &&
   +\frac{q ((-26-6 q+3 q^2) \chi_{1z}+(3-6 q-26 q^2) \chi_{2z}) (M \Omega)}{24 (1+q)^4}\cr &&
   -\frac{(M \Omega)^{4/3}}{5184 (1+q)^6}\big[-4509 \pi ^2 q-27054 \pi ^2 q^3-4509 \pi
   ^2 q^5
+144 (9+12 \chi_{1x}^2+66 \chi_{1y}^2-14 \chi_{1z}^2)
\cr &&
+36 q (1841-144 \chi_{1x}^2
+1128 \chi_{1y}^2
+8 \chi_{1z}^2-432 \chi_{1x} \chi_{2x}+792 \chi_{1y} \chi_{2y}+240
   \chi_{1z} \chi_{2z})
\cr &&
+144 q^6 (9+12 \chi_{2x}^2+66 \chi_{2y}^2-14 \chi_{2z}^2)\cr &&+36 q^5 (1841-432 \chi_{1x} \chi_{2x}-144 \chi_{2x}^2
   +792 \chi_{1y} \chi_{2y}+1128
   \chi_{2y}^2+240 \chi_{1z} \chi_{2z}+8 \chi_{2z}^2)\cr &&+36 q^2 (7256-501 \pi ^2-858 \chi_{1x}^2+1872 \chi_{1y}^2+400 \chi_{1z}^2-1428 \chi_{1x} \chi_{2x}-426 \chi_{2x}^2 +
   3024 \chi_{1y}
   \chi_{2y}+408 \chi_{2y}^2\cr &&+752 \chi_{1z} \chi_{2z}+216 \chi_{2z}^2)+36 q^4 (7256-501 \pi ^2-426 \chi_{1x}^2+408 \chi_{1y}^2+216 \chi_{1z}^2-1428 \chi_{1x} \chi_{2x}-858
   \chi_{2x}^2
   \cr &&
+3024 \chi_{1y} \chi_{2y}+1872 \chi_{2y}^2+752 \chi_{1z} \chi_{2z}+400 \chi_{2z}^2)
\cr &&
+8 q^3 (49043-4914 \chi_{1x}^2+6372 \chi_{1y}^2+2484 \chi_{1z}^2-8964 \chi_{1x}
   \chi_{2x}\cr &&
-4914 \chi_{2x}^2
   +20088 \chi_{1y} \chi_{2y}+6372 \chi_{2y}^2+4608 \chi_{1z} \chi_{2z}+2484 \chi_{2z}^2)\big] 
\cr &&
   +\frac{ (M \Omega)^{5/3}}{288 (1+q)^6}\big[-36 (1+q)^2 (4+3
   q) \chi_{1z}^3-36 q (1+q)^2 (11+10 q) \chi_{1z}^2 \chi_{2z}+\chi_{1z} (144 (3+2 \chi_{1x}^2-\chi_{1y}^2)
\cr &&
   +12 q (163+66 \chi_{1x}^2-33 \chi_{1y}^2+48 \chi_{1x} \chi_{2x}-24
   \chi_{1y} \chi_{2y})+q^2 (3233+720 \chi_{1x}^2-360 \chi_{1y}^2+1584 \chi_{1x} \chi_{2x}
\cr &&+288 \chi_{2x}^2-792 \chi_{1y} \chi_{2y}
   -144 \chi_{2y}^2-360 \chi_{2z}^2)
\cr &&
+9 q^4
   (147+48 \chi_{1x} \chi_{2x}+80 \chi_{2x}^2-24 \chi_{1y} \chi_{2y}-40 \chi_{2y}^2-128 \chi_{2z}^2)
\cr &&
+18 q^3 (159+12 \chi_{1x}^2-6 \chi_{1y}^2+80 \chi_{1x} \chi_{2x}
   +44\chi_{2x}^2-40 \chi_{1y} \chi_{2y}-22 \chi_{2y}^2-62 \chi_{2z}^2)
\cr &&+18 q^5 (15+12 \chi_{2x}^2-6 \chi_{2y}^2-22 \chi_{2z}^2))+q \chi_{2z} (54 (5+4 \chi_{1x}^2-2
   \chi_{1y}^2)
\cr &&
   +9 q (147+80 \chi_{1x}^2-40 \chi_{1y}^2+48 \chi_{1x} \chi_{2x}-24 \chi_{1y} \chi_{2y})
\cr &&
+q^3 (3233+288 \chi_{1x}^2-144 \chi_{1y}^2+1584 \chi_{1x} \chi_{2x}+720
   \chi_{2x}^2
   -792 \chi_{1y} \chi_{2y}-360 \chi_{2y}^2-360 \chi_{2z}^2)
\cr &&
+12 q^4 (163+48 \chi_{1x} \chi_{2x}+66 \chi_{2x}^2-24 \chi_{1y} \chi_{2y}-33 \chi_{2y}^2-33
   \chi_{2z}^2)
\cr &&+18 q^2 (159+44 \chi_{1x}^2
   -22 \chi_{1y}^2+80 \chi_{1x} \chi_{2x}+12 \chi_{2x}^2-40 \chi_{1y} \chi_{2y}-6 \chi_{2y}^2-6 \chi_{2z}^2)
\cr &&+144 q^5 (3+2
   \chi_{2x}^2-\chi_{2y}^2-\chi_{2z}^2))\big] ,
\eea
\end{widetext}
for the ADMTT coordinate separation of the holes, $r$.

From these expressions, we can compute the relevant QC information
of the BBH, given an initial orbital frequency. In what follows, we
will use this information to compute the precise location of the individual holes
with respect to the 3.5PN center of mass to optimize the initial numerical grids
set up in the subsequent simulations, and also evaluate
the instantaneous inwards momentum of the holes via radiation reaction, as
in Ref.~\cite{Healy:2017zqj} it
was observed to produce even smaller measured eccentricities in the
actual full numerical simulations.

\subsection{Center of mass vector}\label{sec:CofM}

To conveniently specify locations of the two black holes, we need
to compute the
center of mass vector $\vec{G}$ at 3.5PN order. The orbital components
of the center of mass vector are derived in Ref.~\cite{Damour:2000kk},
the leading-order (LO) and next-to-leading order (NLO) spin-orbit (SO) part from
Ref.~\cite{Damour:2007nc},
the $S_1$-$S_2$ interaction term from Ref.~\cite{Steinhoff:2008zr},
the $S_1^2$ (and $S_2^2$) part from Ref.~\cite{Hergt:2008jn}
and finally the next-to-next-to-leading order
(NNLO) SO component from Ref.~\cite{Hartung:2011te}.
Imposing $\vec{G} = 0$ as well as $x_1-x_2 = r$, $y_1=y_2$ and $z_1=z_2$, we find
\begin{widetext}
\bea
x_1 &=& \frac{q r}{1+q}+\frac{1}{16 (1+q)^7}\bigg[8 M (-1+q) q (1+q)^4-\frac{8 P^2 (-1+q) (1+q)^8 r}{M^2 q}\bigg]
\cr && +\frac{1}{16 (1+q)^7}\bigg[\frac{4 P^2 (-1+q) (1+q)^6 (5+8 q+5 q^2)}{M q}
\cr && -\frac{4
   M^2 (-1+q) q (1+q)^2 (1+q^2)}{r}+\frac{2 P^4 (-1+q) (1+q)^{12} r}{M^4 q^3}\bigg]+\frac{P (-\chi_{1z}+q \chi_{2z})}{2
   (1+q)}\cr && +\frac{1}{16 (1+q)^7}\bigg[\frac{2 P^3 (1+q)^8 ((2+q) \chi_{1z}-(1+2 q) \chi_{2z})}{M^2 q}+\frac{8 M P (1+q)^4 ((5+3 q) \chi_{1z}-q^2 (3+5 q)
   \chi_{2z})}{r}\bigg]
\cr && 
   +\frac{1}{16 (1+q)^7}\bigg[\frac{P^5 (1+q)^{10} (-((1+q (1+q)^2) \chi_{1z})+\chi_{2z}+q
   (2+q+q^2) \chi_{2z})}{M^4 q^3}
\cr && +\frac{P^3 (1+q)^6 ((-45-42 q-2 q^2+18 q^3) \chi_{1z}+(-18+2 q+42 q^2+45
   q^3) \chi_{2z})}{M q r}
\cr && +\frac{M^2 P (1+q)^2 }{r^2}((-79+q (-75+q (117+q (187+50 q)))) \chi_{1z}
\cr && +q (-50+q (-187+q (-117+q (75+79 q))))
   \chi_{2z})
\bigg]
\cr && +\frac{1}{16 (1+q)^7}\bigg[-\frac{9 P^4 (1+q)^{10} (-1+q^3)}{M^3 q^3}+\frac{P^2 (-1+q) (1+q)^4 (-30-125 q-198 q^2-125 q^3-30
   q^4)}{q r}
\cr && -\frac{P^6 (-1+q) (1+q)^{14} (1+q^2) r}{M^6 q^5} -\frac{1}{r^2}\Big(2 M^3 q (1-12 \chi_{1x}^2+10 \chi_{1y}^2+10 \chi_{1z}^2+q (2+26 \chi_{1y}^2+26 \chi_{1z}^2
\cr && -4 \chi_{1x} (7 \chi_{1x}+2 \chi_{2x})+4 \chi_{1y} \chi_{2y}+ 4
   \chi_{1z} \chi_{2z}+q^4 (-1+12 \chi_{2x}^2-10 \chi_{2y}^2-10 \chi_{2z}^2)
\cr &&
+q (5-20 \chi_{1x}^2 +22\chi_{1y}^2+22 \chi_{1z}^2-8 \chi_{1x} \chi_{2x}+4 \chi_{2x}^2+4 \chi_{1y} \chi_{2y}-6 \chi_{2y}^2+4
   \chi_{1z} \chi_{2z}-6 \chi_{2z}^2)
\cr && -q^2 (5+4 \chi_{1x}^2-6 \chi_{1y}^2-6 \chi_{1z}^2-8 \chi_{1x}
   \chi_{2x}-20 \chi_{2x}^2
\cr && +4 \chi_{1y} \chi_{2y}+22 \chi_{2y}^2+4 \chi_{1z} \chi_{2z}+22 \chi_{2z}^2)+q^3 (4 \chi_{2x} (2 \chi_{1x}+7 \chi_{2x})
\cr && -2 (1+2 \chi_{1y} \chi_{2y}+13 \chi_{2y}^2+2 \chi_{1z} \chi_{2z}+13 \chi_{2z}^2))))\Big)\bigg] ,
\\
y_1 &=& \frac{M^3 q (\chi_{1x} (4 \chi_{1y}+q \chi_{2y})-q \chi_{2x} (\chi_{1y}+4 q \chi_{2y}))}{2 (1+q)^4 r^2} ,
\\
z_1 &=& \frac{P (\chi_{1x}-q \chi_{2x})}{2 (1+q)}+\frac{1}{16 M^4 q^3 (1+q)^5 r^2}\bigg[-2 M^2 P^3 q^2 (1+q)^6 r^2 ((2+q) \chi_{1x}-(1+2 q) \chi_{2x})\cr && -4 M^5 P q^3 (1+q)^2 r
   ((6+q (9+5 q)) \chi_{1x}-q (5+9 q+6 q^2) \chi_{2x})\bigg]\cr && +\frac{1}{16 M^4 q^3 (1+q)^5 r^2}\bigg[P^5 (1+q)^8 r^2 ((1+q
   (1+q)^2) \chi_{1x}-(1+q (2+q+q^2)) \chi_{2x})\cr && -M^3 P^3 q^2 (1+q)^4 r ((-32+q (-50+q (-11+9 q))) \chi_{1x}+(-9+q (11+50 q+32 q^2)) \chi_{2x})\cr && +M^6 P q^3 ((47-3 q (-17+q (19+q (33+10 q)))) \chi_{1x}+q (30+q (99+q (57-q (51+47
   q)))) \chi_{2x})\bigg]\cr && +\frac{M^3 q (\chi_{1x} (4 \chi_{1z}+q \chi_{2z})-q \chi_{2x} (\chi_{1z}+4 q
   \chi_{2z}))}{2 (1+q)^4 r^2} .
\eea
\end{widetext}
Note that here we use the variables
$r=r(M\Omega)$ and $P=P_t(M\Omega)=P_1^y=-P_2^y$ in the previous subsection.

\subsection{Radiation reaction}

So far, all the parameters were calculated using the conservative part of the
3.5PN Hamiltonian. To take into account the radial momentum from the
energy radiated by the binary, we also have to consider the energy flux and
the mass rate change as follows.
\begin{equation}\label{drdt}
\frac{dr}{dt} = \left. -\left(\frac{dE_{\rm GW}}{dt} + \dot{M}\right)
\right/ \left(\frac{dE_{\rm Orb}}{dr}\right),
\end{equation}
where the expressions for $\dot{M}$ and $E_{\rm GW}$ for the non-precessing case are given in Ref.~\cite{Brown:2007jx}, while for those in the precessing case we can use Ref.~\cite{Ossokine:2015vda}.

Now, from the Hamilton equation
\begin{equation}
\dot{r} = \frac{\partial H}{\partial P_r},
\end{equation}
excluding higher $P_r$ order terms, we derive the following relationship
between $\dot{r}$ and $P_r$,
\begin{equation}\label{Pr definition}
P_r = \dot{r}\left[\lim_{P_r\to 0}\frac{P_r}{\partial H/\partial P_r}\right].
\end{equation}

This leads to
\begin{widetext}
\bea
\frac{dr}{dt}&=&\frac{M^{7/2}}{r^{7/2}}\Bigg[\bigg(-\frac{13 q^2}{4 (q+1)^4}-\frac{3 q}{2 (q+1)^4}\bigg) \chi_{1y} \chi_{2x}+\bigg(-\frac{3 q^3}{2 (q+1)^4}-\frac{q^2}{4 (q+1)^4}\bigg) \chi_{2y} \chi_{2x}
\cr &&+ \bigg(-\frac{q^2}{4 (q+1)^4}-\frac{3 q}{2 (q+1)^4}\bigg) \chi_{1x} \chi_{1y}+\bigg(-\frac{3 q^3}{2 (q+1)^4}-\frac{13 q^2}{4 (q+1)^4}\bigg) \chi_{1x} \chi_{2y}\Bigg]
\cr &&
   +\frac{(1+q)^2 P_r}{qM} \Bigg\{1- \left(\frac{M}{r}\right) \frac{7 q^2+15 q+7}{2
   (q+1)^2}
   +\left(\frac{M}{r}\right)^2\frac{47 q^4+229 q^3+363 q^2+229 q+47}{8 (q+1)^4}
\cr &&
   +\left(\frac{M}{r}\right)^{5/2}\bigg[\bigg(\frac{q^2}{(q+1)^3}+\frac{11 q}{4 (q+1)^3}+\frac{3}{(q+1)^3}\bigg) \chi_{1z}+\bigg(\frac{3 q^3}{(q+1)^3}+\frac{11 q^2}{4 (q+1)^3}+\frac{q}{(q+1)^3}\bigg)
   \chi_{2z}\bigg]
\cr &&+\left(\frac{M}{r}\right)^3\bigg[-\frac{121 q^6}{16 (q+1)^6}-\frac{\pi ^2 q^5}{16 (q+1)^6}-\frac{163 q^5}{3 (q+1)^6}-\frac{\pi ^2 q^4}{4 (q+1)^6}-\frac{1831 q^4}{12 (q+1)^6}-\frac{3 \pi ^2 q^3}{8
   (q+1)^6}
\cr &&-\frac{3387 q^3}{16 (q+1)^6}-\frac{\pi ^2 q^2}{4 (q+1)^6}-\frac{1831 q^2}{12 (q+1)^6}-\frac{\pi ^2 q}{16 (q+1)^6}-\frac{163 q}{3 (q+1)^6}
\cr &&
   +\bigg(\frac{q^4}{2 (q+1)^6}+\frac{11 q^3}{2
   (q+1)^6}+\frac{14 q^2}{(q+1)^6}+\frac{27 q}{2 (q+1)^6}+\frac{9}{2 (q+1)^6}\bigg) \chi_{1x}^2
\cr &&
   +\bigg(-\frac{9 q^3}{4 (q+1)^6}-\frac{27 q^2}{4 (q+1)^6}-\frac{27 q}{4 (q+1)^6}-\frac{9}{4 (q+1)^6}\bigg)
   \chi_{1y}^2
\cr &&+\bigg(-\frac{9 q^3}{4 (q+1)^6}-\frac{27 q^2}{4 (q+1)^6}-\frac{27 q}{4 (q+1)^6}-\frac{9}{4 (q+1)^6}\bigg) \chi_{1z}^2
\cr &&
   +\bigg(\frac{9 q^6}{2 (q+1)^6}+\frac{27 q^5}{2 (q+1)^6}+\frac{14
   q^4}{(q+1)^6}+\frac{11 q^3}{2 (q+1)^6}+\frac{q^2}{2 (q+1)^6}\bigg) \chi_{2x}^2
\cr &&+\bigg(-\frac{9 q^6}{4 (q+1)^6}-\frac{27 q^5}{4 (q+1)^6}-\frac{27 q^4}{4 (q+1)^6}-\frac{9 q^3}{4 (q+1)^6}\bigg)
   \chi_{2y}^2
\cr &&+\bigg(-\frac{9 q^6}{4 (q+1)^6}-\frac{27 q^5}{4 (q+1)^6}-\frac{27 q^4}{4 (q+1)^6}-\frac{9 q^3}{4 (q+1)^6}\bigg) \chi_{2z}^2
\cr &&
   +\bigg(\frac{3 q^5}{(q+1)^6}+\frac{11 q^4}{(q+1)^6}+\frac{16
   q^3}{(q+1)^6}+\frac{11 q^2}{(q+1)^6}+\frac{3 q}{(q+1)^6}\bigg) \chi_{1x} \chi_{2x}\cr &&
   +\bigg(\frac{3 q^5}{2 (q+1)^6}+\frac{7 q^4}{(q+1)^6}+\frac{11 q^3}{(q+1)^6}+\frac{7 q^2}{(q+1)^6}+\frac{3 q}{2
   (q+1)^6}\bigg) \chi_{1y} \chi_{2y}
\cr &&
   +\bigg(\frac{3 q^5}{2 (q+1)^6}+\frac{7 q^4}{(q+1)^6}+\frac{11 q^3}{(q+1)^6}+\frac{7 q^2}{(q+1)^6}+\frac{3 q}{2 (q+1)^6}\bigg) \chi_{1z} \chi_{2z}-\frac{121}{16
   (q+1)^6}\bigg]
\cr && +\left(\frac{M}{r}\right)^{7/2}\bigg[\bigg(-\frac{53 q^5}{16 (q+1)^6}-\frac{357 q^4}{16 (q+1)^6}-\frac{1097 q^3}{16 (q+1)^6}-\frac{743 q^2}{8 (q+1)^6}-\frac{421 q}{8 (q+1)^6}-\frac{9}{(q+1)^6}\bigg)\chi_{1z}
\cr && + \bigg(-\frac{9 q^6}{(q+1)^6}-\frac{421 q^5}{8 (q+1)^6}-\frac{743 q^4}{8 (q+1)^6}-\frac{1097 q^3}{16 (q+1)^6}-\frac{357 q^2}{16 (q+1)^6}-\frac{53 q}{16 (q+1)^6}\bigg) \chi_{2z}\bigg]
\Bigg\} .
\label{rdotofPr}
\eea
\end{widetext}

Finally, using Eq.~\eqref{drdt}, we can invert Eq.~\eqref{rdotofPr} and obtain $P_r$
in terms of $r$. Given our particular initial configuration of the
black holes along the $x$-axis, we have $P_r = P_{1x} = -P_{2x}$. 

While this consistently completes our 3.5PN determination of the QC
parameters for numerical simulations of BBHs, we would like to
consider the special case of intermediate to small mass ratio binaries to
extend and improve the reduction of eccentricity methods by considering for
the large hole as a Schwarzschild or Kerr background
(as a sort of resummation of PN terms)
for the linear dependence with the mass ratio $q$,
while preserving the 3.5PN terms at all higher powers of $q$.

\subsection{Small mass ratio}\label{sec:Kerr}

In this subsection, we discuss an additional method, particularly useful to treat
small mass ratio systems,
in which we combine the particle limit and the 3.5PN treatment of the problem.
In particular, starting from the Kerr metric $g^{\rm Kerr}_{\mu\nu}$ written in
quasi-isotropic (QISO) coordinates, we have
\bea\label{KHamiltonian}
g^{\mu\nu}_{\rm Kerr}p_\mu p_\nu 
&=& E^2 g^{tt} + P_{\phi}^2 g^{\phi \phi} + 2 g^{\phi t} E P_{\phi} 
\cr &=& -1 .
\eea

This expression gives us a formula to compute $E$ which is our particle's Hamiltonian in QISO coordinates. To verify that this is the correct Hamiltonian, one can use the transformation of coordinates (see Ref.~\cite{Hergt:2007ha})
\bea\label{ADMtoqiso}
x_{\rm QISO}^k &=& x_{\rm ADM}^k+\xi^k
\cr &=& x_{\rm ADM}^k-\frac{1}{4}\frac{M^2a^2 n^k}{r_{\rm ADM}}-\frac{7}{16}\frac{M^4a^2n^k}{r_{\rm ADM}^3},
\eea
where $a = S^z/M^2$ 
is the (nondimensional) spin of the more massive black hole, then perform an expansion of $E$
obtained from Eq.~\eqref{KHamiltonian} and verify that it reproduces the Hamiltonian
in Eq.~\eqref{Hamiltonian}.

Using this method, the particle Hamiltonian in QISO coordinates
(labeled simply by $r=r_{\rm QISO}/M$ for the sake of simplicity) becomes
\begin{widetext}
\bea
H_{\rm Kerr} &=& -\frac{64 r^3}{(a^2-4 r^2-4 r-1)^3+16 a^2 r^2 (a^2-4 r (r+3)-1)}\cr &&
\times \Bigg[-\frac{a^2 (a-2 r-1) (a+2 r+1) \sqrt{\frac{4 (L^2+4)}{a^2+4 r^2-1}-\frac{4
   (a^2-1) (L^2+4)}{(a^2+4 r^2-1)^2}+\frac{2 a+2}{a^2+2 a r+a}-\frac{2 (a-1)}{a
   (a-2 r-1)}+1}}{4 r}\cr &&
   +\frac{(a-2 r-1) (a-2 r+1) (a+2 r-1) (a+2 r+1) \sqrt{(2 r+1)^2-a^2}}{64 r^3}\cr &&
   \times \frac{1}{(a^2+4 r^2-1)^2}\bigg(-a^6+a^4 (3-4 (r-3)
   r)+a^2 (8 r (r (-2L^2+2 r (r+6)-7)-3)-3)\cr &&
   +(2 r+1)^2 (8 r (r (2
  L^2+2 r (r+2)+3)+1)+1)\bigg)^{1/2}-2 a
  L\Bigg] ,
\label{HKerr}
\eea
\end{widetext}
where $L=L^z/(M^2\eta)$ is the orbital angular momentum of the system. We point out that here,
in the particle limit, we are assuming that the particle orbits in the equatorial plane.

Using the Hamiltonian in Eq.~\eqref{HKerr} and the method described in
Sec.~\ref{3.5PN Hamiltonian}, one can derive the initial parameters
in the particle limit $\eta\to0$. The idea is now to combine the 3.5PN
Hamiltonian with the particle limit Hamiltonian. To do so, we
should first specify how a vector transforms under the coordinate
transformation in Eq.~\eqref{ADMtoqiso},
\bea
V^\mu_{\rm QISO}(x^\sigma_{\rm QISO}) &=& V^\nu_{\rm ADM}(x^\sigma_{\rm ADM})\frac{\partial x^\mu_{\rm ADM}}{\partial x^\nu_{\rm QISO}}
\cr &=& V^\nu_{\rm ADM}(x^\sigma_{\rm ADM})(\delta^\mu_\nu+\xi^\mu,_\nu)
\cr &=& V^\mu_{\rm ADM}(x^\sigma_{\rm ADM})+V^\nu_{\rm ADM}(x^\sigma_{\rm ADM}) \xi^\mu,_\nu ,
\cr &&
\eea
relating the QISO coordinates to our preferred ADMTT gauge.
We note that only the $r$-component of a vector will change under this
coordinate transformation in Eq.~\eqref{ADMtoqiso}.

For example, to compute the angular momentum $P_\phi$, we have
\begin{equation}
P^{\phi} = P^{\phi}_{\rm 3.5PN} - P^{\phi}_{\rm 3.5PN}\big|_{\eta=0} + P^{\phi}_{\rm Kerr} .
\end{equation}
Similarly, we have
\begin{equation}
M^{\rm ADM} = M^{\rm ADM}_{\rm 3.5PN} - M^{\rm ADM}_{\rm 3.5PN}\big|_{\eta=0} + M^{\rm ADM}_{\rm Kerr} ,
\end{equation}
and
\begin{equation}
r^{\rm QISO} = r^{\rm QISO}_{\rm 3.5PN} - r^{\rm QISO}_{\rm 3.5PN}\big|_{\eta=0} + r^{\rm QISO}_{\rm Kerr} .
\end{equation}

This distinction will be particularly important in the next section where we
put to test this choice of the Kerr QISO coordinates to associate with
the numerical coordinates versus the standard use of the ADMTT coordinates.

\section{Full Numerical Techniques and Results}\label{sec:FN}

To test these theoretically improved choices of QC parameters,
we perform a set of comparative full numerical simulations for configurations
of current interest in the intermediated mass ratio regime. We will consider
three basic configurations all bearing  mass ratio $q=1/16$ and spins
of the larger black hole aligned or counteraligned with the orbital angular
momentum of the system with intrinsic values $\chi=-0.8$, $-0.4$, and $+0.8$,
while the smaller hole is nonspinning.
Initial separations of the binary have been chosen decreasingly,
$d/M=10.0$, $9.5$, and $8.0$, to compensate in part for the hangup/hangdown effect 
\cite{Campanelli:2006uy,Healy:2018swt} that prompts or delays merger
for aligned or counteraligned spin-orbit coupling, respectively.

In Table~\ref{tab:tmerger}, we report the time to merger for each of
the three spinning configurations while choosing five different methods
to determine QC orbits. We identify the initial data by
the methods as follows: id 0 is the 3PN  order given in
Ref.~\cite{Healy:2017zqj}, id 1 is the improvement to 3.5PN order given
here, id 2 is the partial resummation of the nonspinning terms into
a Schwarzschild background for the linear in $q$ terms, id 3 is
the use of the Kerr metric in the QISO coordinates,
and id 4 is the direct use of the ADMTT gauge transformation
given in Eq.~\eqref{ADMtoqiso} to express also the Kerr metric
in these ADMTT coordinates.


\begin{table}
  \centering
  \caption{Number of orbits and time to merger
 \label{tab:tmerger}}
 \begin{tabular}{rllll}
\hline
spin & id & ID Name &orbits & $t_{\rm merg}$ \\
\hline
$-0.8$ & 0 & 3PN & 6.57 & 1118 \\
$-0.8$ & 1 & 3.5PN &6.93 & 1196 \\
$-0.8$ & 2 & 3.5PN+Schwarzschild & 7.04 & 1220 \\
$-0.8$ & 3 & 3.5PN+KerrQISO &7.14 & 1241 \\
$-0.8$ & 4 & 3.5PN+KerrADMTT &7.41 & 1302 \\
\hline
$-0.4$ & 0 & 3PN &9.31 & 1476 \\
$-0.4$ & 1 & 3.5PN &9.48 & 1513 \\
$-0.4$ & 2 & 3.5PN+Schwarzschild &9.60 & 1538 \\
$-0.4$ & 3 & 3.5PN+KerrQISO &9.66 & 1551 \\
$-0.4$ & 4 & 3.5PN+KerrADMTT &9.72 & 1564 \\
\hline
$+0.8$ & 0 & 3PN &22.74 & 2591 \\
$+0.8$ & 1 & 3.5PN &22.09 & 2477 \\
$+0.8$ & 2 & 3.5PN+Schwarzschild &22.31 & 2515 \\
$+0.8$ & 3 & 3.5PN+KerrQISO &22.15 & 2488 \\
$+0.8$ & 4 & 3.5PN+KerrADMTT &22.39 & 2528 \\
\hline 
\end{tabular} 
\end{table} 

Notably, the results for the `hangdown' configurations,
with $\chi=-0.8,$ and $\chi=-0.4$ show an increase in the merger times
$t_{\rm merg}$ as well as in the number of orbits to merger with
the additions of term in the QC method, i.e.,
with increasing id number.
This can be interpreted as due to a lowering of the 
eccentricity (as higher eccentricity is more efficient
in radiating gravitational energy \cite{Peters:1964zz}),
and will be confirmed below with direct measures of the
eccentricity in each id case.
The case $\chi=+0.8$ is less clear since the 3PN choice
seems to produce eccentricity by an overshoot of the holes
(at periastron instead of apastron), but it is also to
note here the close starting point at a distance coordinate
of $d/M=8.0$ and the large number of orbits to merger
due to the strong hangup effect.

\begin{figure}[!ht]
\centering
\includegraphics[width=.48\textwidth]{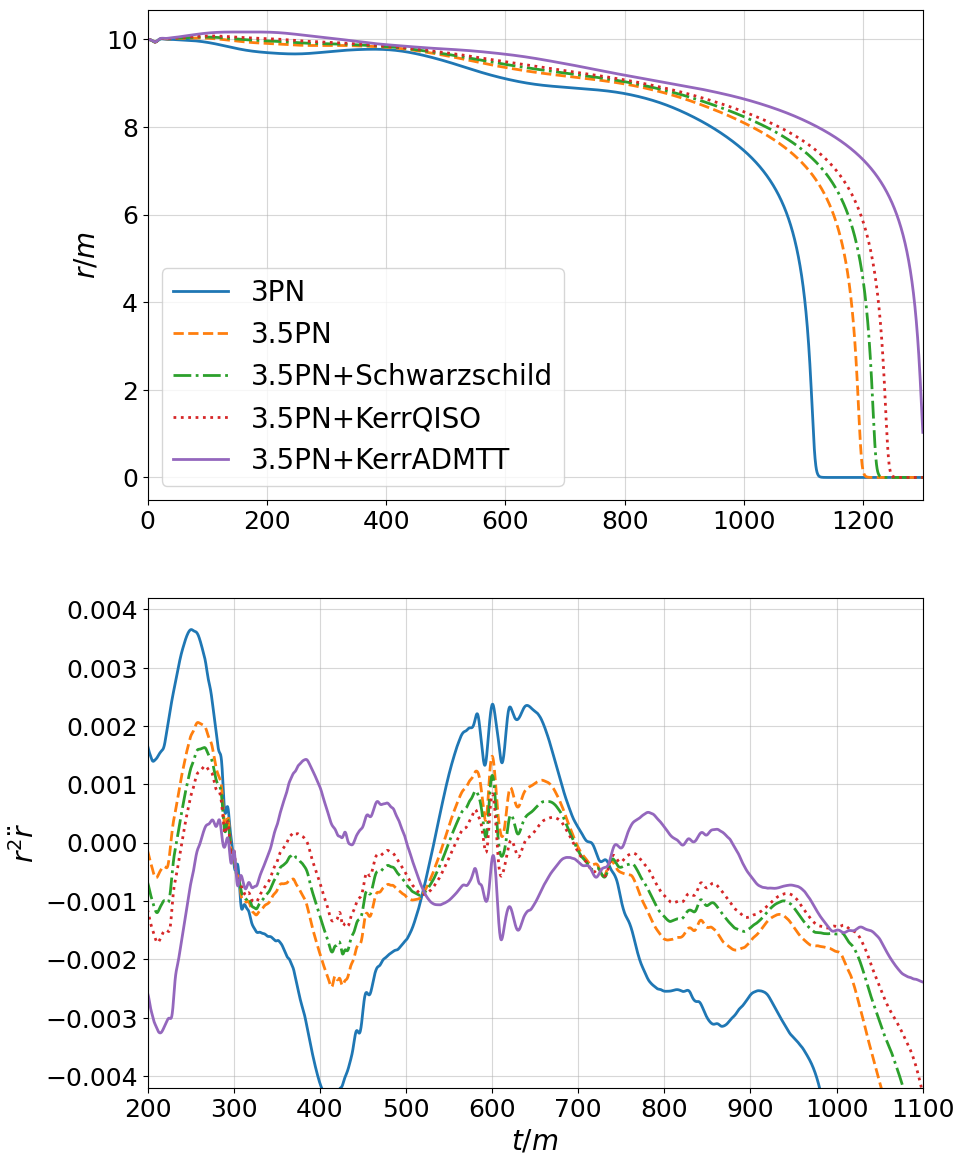}
\caption{Evolution of coordinate distance (top)
and eccentricity measure $r^2\ddot r$ (bottom)
for the $q=1/16, d=10M, \chi=-0.8$ case.
}
\label{fig:spin-0.8_redecc}
\end{figure}

\begin{figure}[!ht]
\centering
\includegraphics[width=.48\textwidth]{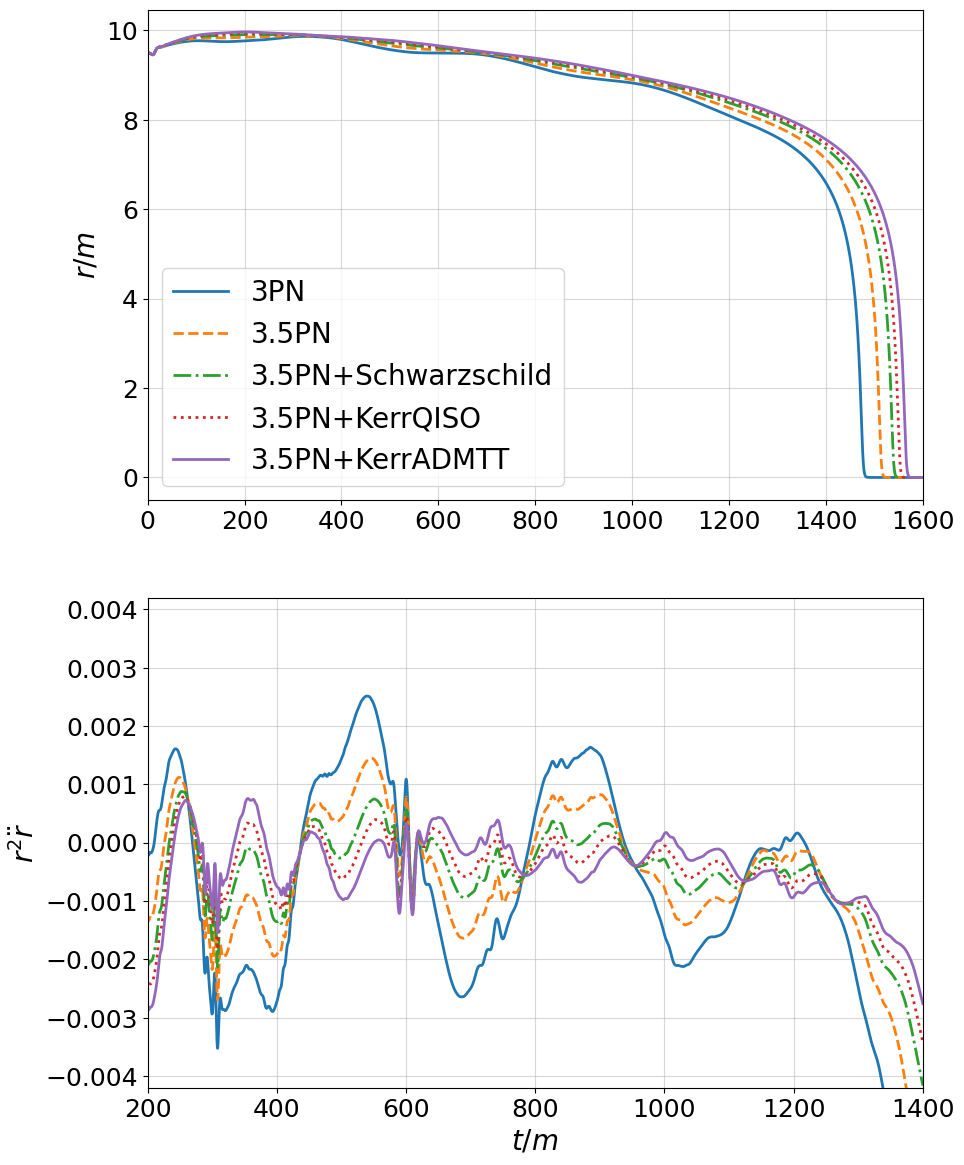}
\caption{Evolution of coordinate distance (top)
and eccentricity measure $r^2\ddot r$ (bottom)
for the $q=1/16, d=9.5M, \chi=-0.4$ case.
}
\label{fig:spin-0.4_redecc}
\end{figure}

\begin{figure}[!ht]
\centering
\includegraphics[width=.48\textwidth]{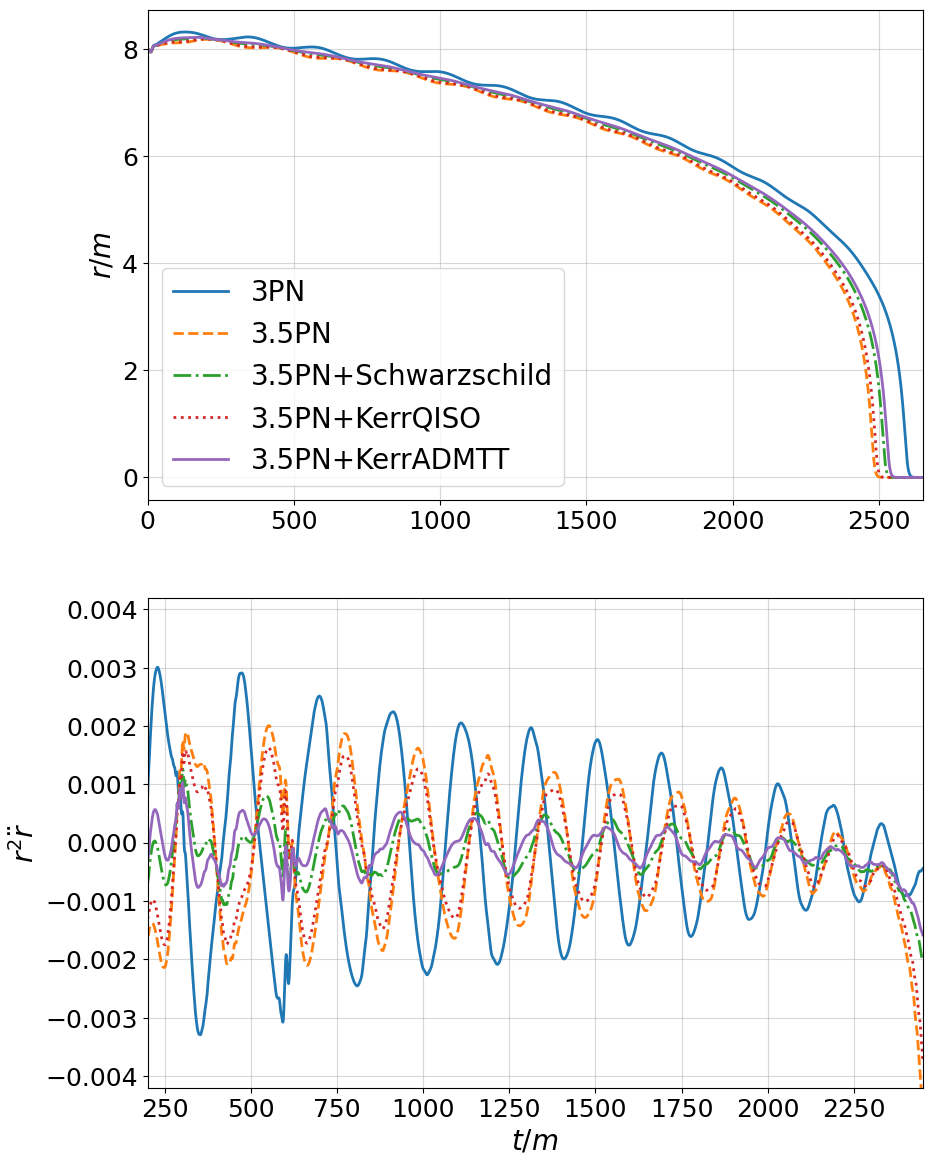}
\caption{Evolution of coordinate distance (top)
and eccentricity measure $r^2\ddot r$ (bottom)
for the $q=1/16, d=8M, \chi=+0.8$ case.
}
\label{fig:spin0.8_redecc}
\end{figure}

We also display here the distance evolution and a measure
of its eccentricity \cite{Campanelli:2008nk},
\bea
e\cos(\Omega t) \sim r^2\ddot r ,
\eea
for each of the three binary configurations studied here in
Figs.~\ref{fig:spin-0.8_redecc}--\ref{fig:spin0.8_redecc}.
Each of those figures clearly show the different merger times
in spite of starting at the same initial separation. The
3PN QC data is clearly the one displaying the
largest eccentricity in coordinates. The amplitude of the
orbital oscillations is more clearly exposed in the bottom plots
of $r^2\ddot r$ and show already a reduction factor of
$\times2$--$3$ for the 3.5PN choice, and further reduction for the
resummation cases. Next we will quantify these eccentricities with 
invariant methods.


To assess the eccentricity reduction more precisely in a
coordinate independent way, we now consider the eccentricity
measured by the amplitude and phase oscillation of $\psi_{4(22)}$, the
Weyl scalar waveform $(\ell,m)=(2,2)$ mode as given in Ref.~\cite{Healy:2017zqj}
given by
\begin{equation}
\psi_{4(22)} = A_{22}(t) \exp [i \varphi_{22}(t)] .
\end{equation}
Hence we define the waveform-based eccentricity measures
\begin{equation}
  e_{A22} = \left(\frac{8}{39}\right){\rm Amp}\left(\frac{A_{22}(t) -
  A_{22 \rm sec}(t)}{A_{22}(t)}\right) \,,
  \label{ecc_eA}
\end{equation}
\begin{equation}
  e_{\omega22} = \left(\frac{8}{21}\right){\rm Amp}\left(\frac{\omega_{22}(t) - \omega_{22 \rm sec}(t)}{
  \omega_{22}(t)}\right) \,,
\label{ecc_ew}
\end{equation}
where $\omega_{22} = d\varphi_{22}/dt$.

The results of these evaluations of the eccentricity
from the initial two orbits amplitude $e_{A22}$ oscillations
and frequency (when possible to measure meaningfully) $e_{\omega22}$,
are given in Table~\ref{tab:eccentricity}, where we also included the
$e_{\rm spd}$ obtained from the using the simple proper distance $D$ along the line
joining the horizons of the holes to evaluate $D^2\ddot D$ oscillations amplitude.
While this different measures provide a possible definition of eccentricity
(which is a Newtonian concept, not necessarily well nor uniquely defined in general relativity),
we see the decreasing pattern with the successive approaches within each of the
eccentricity definitions. For each of those measures, in particular for all
$e_{\rm spd}$, and $e_{A22}$, we see the same relative pattern behavior of decrease of
eccentricity, favoring the Kerr background choice in ADMTT coordinates to be identified
with the full numerical ones.


\begin{table}
  \centering
  \caption{Waveforms eccentricity.
  }
 \label{tab:eccentricity}
 \begin{tabular}{lllll}
\hline
id &    spin &   $e_{\rm spd}$ & $e_{A22}$    & $e_{\omega22}$ \\
\hline
0  & $-0.8$ & 1.82e-2 & 3.04e-2 &  \\
1  & $-0.8$ & 8.37e-3 & 1.03e-2  & \\   
2  & $-0.8$ & 5.61e-3 & 7.37e-3   &  \\
3  & $-0.8$ & 3.29e-3 & 4.53e-3 &    \\
4  & $-0.8$ & 3.30e-3 & 1.90e-3  &   \\
\hline
0  & $-0.4$ & 7.95e-3 & 7.09e-3 & \\
1  & $-0.4$ & 4.05e-3 & 3.52e-3 & \\
2  & $-0.4$ & 1.47e-3 & 1.18e-3 & \\
3  & $-0.4$ & 2.86e-4 & 1.46e-4 & \\
4  & $-0.4$ & 1.39e-3 & 1.48e-3 & \\
\hline
0  & $+0.8$ & 6.00e-3 & 5.72e-3  & 4.77e-3 \\
1  & $+0.8$ & 4.45e-3 & 4.27e-3  & 3.53e-3 \\
2  & $+0.8$ & 1.57e-3 & 1.49e-3  & 1.22e-3 \\
3  & $+0.8$ & 3.50e-3 & 3.36e-3  & 2.77e-3 \\
4  & $+0.8$ & 1.23e-3 & 1.18e-3  & 1.09e-3 \\
\hline 
\end{tabular} 
\end{table} 


To quantify the magnitude of the differences due to eccentricity between the waveforms, we use the matching measure,
\begin{eqnarray}
\mathscr{M} \equiv \frac{\left<h_1\left|\right.h_2\right>}{\sqrt{\left<h_1\left|\right.h_1\right>\left<h_2\left|\right.h_2\right>}},
\end{eqnarray}
as implemented via a complex
overlap as described in Eq.~(2) in Ref.~\cite{Cho:2012ed}:
\begin{eqnarray}
\left< h_1 \left|\right. h_2 \right> & = & 2 \int_{-\infty}^{\infty} \frac{df}{S_n(f)}\left[\tilde{h}_1(f) \tilde{h}_2(f)^* \right],\label{eq:match}
\end{eqnarray} where $\tilde{h}(f)$ is the Fourier transform of $h(t)$ and $S_n(f)$ is the power spectral density of the detector noise (here, taken to be identically equal
to 1 since we are interested in the direct waveforms comparisons).
We adopt the leading modes $(\ell,m)=(2,2)$ of $\psi_4$ for the computations and we do not
maximize over an overall constant time shift and
an overall constant phase shift as is usually done for parameter estimation
of gravitational waves signals~\cite{Lovelace:2016uwp}.

The results are presented in Table~\ref{tab:match} as a cross-match between all
the five approximation of QC parameters for the three cases of spins
aligned and counteraligned studied here. We find the largest deviations between
waveforms from the original 3PN data for the counteraligned spin configurations.
We also note the preference of the ADMTT coordinates in the Kerr resummation case
over the QISO coordinates identification with the numerical coordinates.
The aligned spin case, with its tight orbits due to the strong hangup effect
shows notable agreement between all waveforms.


\begin{table}
  \centering
  \caption{Waveforms cross-matching.
  }
 \label{tab:match}
 \begin{tabular}{l|l|lllll}
         \hline
     & id2  &       0  &        1 &       2  &       3  &       4 \\
\hline
spin & id1  &          &          &          &          &         \\
$-0.8$ 
     & 0    & 1.00000  & 0.71331  & 0.71980  & 0.71410  & 0.70901 \\
     & 1    & 0.71331  & 1.00000  & 0.99664  & 0.95694  & 0.99937 \\
     & 2    & 0.71980  & 0.99664  & 1.00000  & 0.97749  & 0.99311 \\
     & 3    & 0.71410  & 0.95694  & 0.97749  & 1.00000  & 0.94604 \\
     & 4    & 0.70901  & 0.99937  & 0.99311  & 0.94604  & 1.00000 \\
\hline
     & id2  &       0  &       1  &       2  &       3  &       4 \\
\hline
spin & id1  &          &          &          &          &  \\
$-0.4$ 
     & 0    & 1.00000  & 0.97944  & 0.98241  & 0.97955  & 0.95987 \\
     & 1    & 0.97944  & 1.00000  & 0.99988  & 1.00000  & 0.99671 \\
     & 2    & 0.98241  & 0.99988  & 1.00000  & 0.99989  & 0.99536 \\
     & 3    & 0.97955  & 1.00000  & 0.99989  & 1.00000  & 0.99667 \\
     & 4    & 0.95987  & 0.99671  & 0.99536  & 0.99667  & 1.00000 \\
\hline
     & id2  &       0  &       1  &       2  &       3  &       4 \\
\hline
spin & id1  &          &          &          &          &         \\    
$+0.8$ 
     & 0    & 1.00000  & 0.99939  & 0.99976  & 0.99976  & 0.99925 \\
     & 1    & 0.99939  & 1.00000  & 0.99978  & 0.99951  & 0.99902 \\
     & 2    & 0.99976  & 0.99978  & 1.00000  & 0.99994  & 0.99939 \\
     & 3    & 0.99976  & 0.99951  & 0.99994  & 1.00000  & 0.99940 \\
     & 4    & 0.99925  & 0.99902  & 0.99939  & 0.99940  & 1.00000 \\
         \hline 
\end{tabular} 
\end{table} 


In summary, all these results provide a consistent relative reduction of
the eccentricity with the successive PN approximations to
QC orbits and, notably, a preferences of the
ADMTT gauge over the QISO in the Kerr case
(for Schwarzschild isotropic coordinates are also in the
ADMTT gauge) for our full numerical set up.

\begin{figure}[!ht]
\centering
\includegraphics[width=.48\textwidth]{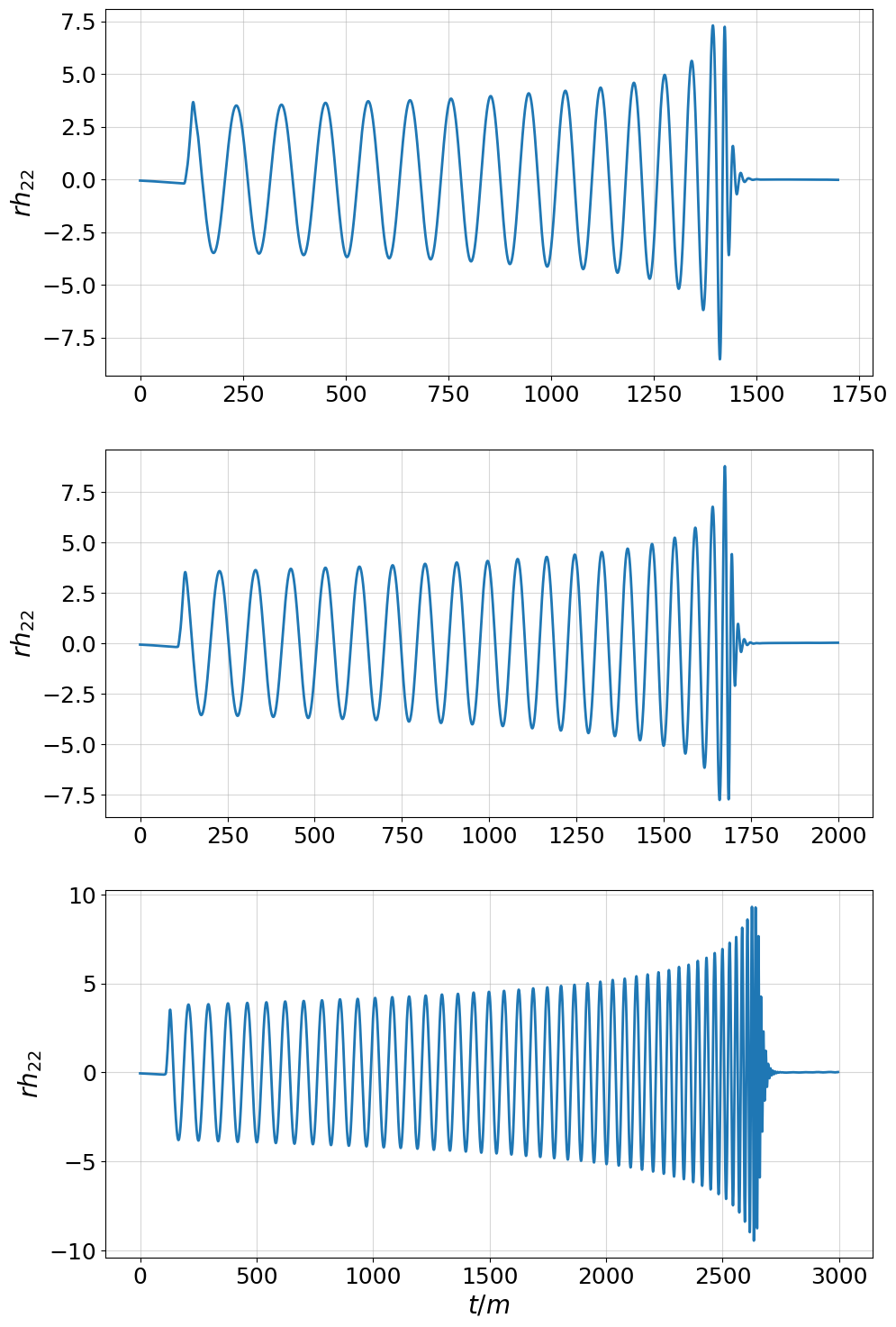}
\caption{Strain waveform for the 3.5PN+KerrADMTT QC data for
  the three cases studied here
  $q=1/16, d=10M, \chi=-0.8$ (top);
  $q=1/16, d=9.5M, \chi=-0.4$ (middle);
  $q=1/16, d=8M, \chi=+0.8$ (bottom); respectively.
}
\label{fig:3waveforms.png}
\end{figure}

We finally construct the strain $h$ waveforms from the double integration
of the Weyl scalar $\psi_4$. The strain was computed by taking the
Fourier transform of $\psi_4$, removing modes in a small region around
$\omega=0$, then dividing by $-\omega^2$ and taking the inverse Fourier
transform as detailed in Refs.~\cite{Campanelli:2008nk,Reisswig:2010di}.
In Fig.~\ref{fig:3waveforms.png}, we display for all three spin cases
the strain $(2,2)$-mode at an observer location $r=113M$
for the QC choice of
the 3.5PN+Kerr method in ADMTT coordinates. Note the smooth amplitude increase
(after initial data settling) up to the merger characteristic of the non-eccentric
(nonprecessing) black hole binaries. These prototypical waveforms will serve
as a base for the construction of a first catalog of full numerical simulations
in the intermediate mass ratio regime.


\section{Conclusions and Discussion}\label{sec:Discussion}

We thus conclude through the analysis of different eccentricity
measures that 3.5PN extension to QC
parameters provides a clear benefit over 3PN for the
close initial configurations needed when dealing with
intermediate to small mass ratio binaries. We also have seen a further
improvement when including the Schwarzschild limit and
when spins are large the inclusion of the Kerr background
in the ADMTT gauge expansion.
These improvements on the low eccentricity initial parameters of
intermediate to small mass ratio binaries will have a direct applications
to the next generation of full numerical simulations catalogues to cover the
parameter space relevant to ground third generation and space LISA gravitational wave observatories.

It is also worth noting here that the parameter space of the simulations
should also include specific coverage of the initial eccentricity, in addition 
to the QC cases, since for some 3G and LISA sources it is expected that some residual
eccentricity of the merging binary may survive while entering their sensitivity low frequency band.
For instance, this can be achieved by choosing initial $e=0.1$, $0.2$, etc. values
from the inception of the full numerical simulations by using Ref.~\cite{Ciarfella:2022hfy}.
An especially intriguing case is GW190521, which matches well to 
highly eccentric comparable masses full numerical simulations
\cite{Gayathri:2020coq}.

\begin{acknowledgments}
The authors thank Giuseppe Ficarra, and Yosef Zlochower
for useful discussions. The authors also gratefully acknowledge
the National Science Foundation (NSF) for financial support from Grant
No.\ PHY-2207920.  Computational resources were also
provided by the Blue Sky, Green Prairies, and White
Lagoon clusters at the CCRG-Rochester Institute of Technology, which
were supported by NSF grants No.\ AST-1028087, No.\ PHY-1229173,
No.\ PHY-1726215, and No.\ PHY-2018420.  This work used the ACCESS
allocation TG-PHY060027N, funded by NSF, and project PHY20007 Frontera,
an NSF-funded Petascale computing system at the Texas Advanced Computing Center.
HN is supported by Grants-in-Aid for Scientific Research, 
No.\ 21H01082, No.\ 21K03582 and No.\ 23K03432 
from the Japan Society for the Promotion of Science (JSPS).
\end{acknowledgments}

\appendix

\section{Approximate initial data in terms of radial distance}

In many practical applications, one chooses to compare full numerical simulations
from the same initial separation of the holes $r$. Here, we provide the explicit
basic expressions for the initial QC parameters $P_t$ and $M_{\rm ADM}$,
as well as the inverse to the Eq.~\eqref{eq:r} for $M\Omega(r)$.

The tangential linear momentum $P_t$ expressed in term of the separation of the holes $r$ in ADMTT coordinates is given by
\begin{widetext}
\bea
\frac{P_t}{M} &=& \frac{1}{128} \sqrt{\frac{M}{r}} \Bigg\{128+\frac{256 M}{r}-\frac{96 ((4+3 q) \chi_{1z}+q (3+4 q) \chi_{2z})}{(1+q)^2}\left(\frac{M}{r}\right)^{3/2}
\cr && +\frac{8 }{(1+q)^2}
\bigg[6 (7-4 \chi_{1x}^2+2 \chi_{1y}^2+2 \chi_{1z}^2)+q (41-48 \chi_{1x}
   \chi_{2x}+24 \chi_{1y} \chi_{2y}+24 \chi_{1z} \chi_{2z}
\cr &&
+6 q (7-4 \chi_{2x}^2+2 \chi_{2y}^2+2 \chi_{2z}^2))\bigg]
\left(\frac{M}{r}\right)^{2}
\cr && -\frac{8 ((72+q (116+q (60+13 q))) \chi_{1z}+q (13+4 q (15+q (29+18 q)))
   \chi_{2z})}{(1+q)^4}\left(\frac{M}{r}\right)^{5/2}
\cr && +\frac{1}{(1+q)^4} \bigg[163 \pi ^2 q (1+q)^2+32 (15-20 \chi_{1x}^2-2 \chi_{1y}^2+16 \chi_{1z}^2)+4 q (-659+56 \chi_{1z}^2+48 \chi_{1x} \chi_{2x}
\cr && -8 \chi_{1y} (20 \chi_{1y}+27 \chi_{2y})-8 q^3 (-15+20
   \chi_{2x}^2+2 \chi_{2y}^2)+q^2 (-659+48 \chi_{1x} \chi_{2x}-8 \chi_{2y} (27 \chi_{1y}+20 \chi_{2y}))
\cr && +q (-1532+118 \chi_{1x}^2-27 \chi_{1z}^2+140 \chi_{1x} \chi_{2x}+118 \chi_{2x}^2-16 (8 \chi_{1y}^2+29 \chi_{1y}
   \chi_{2y}+8 \chi_{2y}^2))
\cr && +2 (60+q (133+60 q)) \chi_{1z} \chi_{2z}+q (-27+8 q (7+16 q)) \chi_{2z}^2)\bigg]\left(\frac{M}{r}\right)^{3}
\cr && +\frac{4 }{(1+q)^6}\bigg[-20 (1+q)^2 (4+q) \chi_{1z}^3-60 q (1+q)^2 (3+2 q)
   \chi_{1z}^2 \chi_{2z}+\chi_{1z} (4 (-87+80 \chi_{1x}^2-20 \chi_{1y}^2)\cr &&+10 q (-67-18 \chi_{1y}^2+36 \chi_{1x} (2 \chi_{1x}+\chi_{2x}))+12 q^2 (-21+40 \chi_{1x}^2+70 \chi_{1x} \chi_{2x}-30 \chi_{2x}^2
\cr &&
-10 (\chi_{1y}-2 \chi_{2y}) (\chi_{1y}+\chi_{2y})+10 \chi_{2z}^2)+5 q^4 (-29+24 (\chi_{1x}-9 \chi_{2x}) \chi_{2x}
\cr && +24 \chi_{2y} (\chi_{1y}+5 \chi_{2y})+96 \chi_{2z}^2)+q^5 (-103-360 \chi_{2x}^2+180 \chi_{2y}^2+180
   \chi_{2z}^2)
\cr && +q^3 (27-20 \chi_{1y}^2+40 (2 \chi_{1x}-3 \chi_{2x}) (\chi_{1x}+9 \chi_{2x})+240 \chi_{1y} \chi_{2y}+660 \chi_{2y}^2+420 \chi_{2z}^2))
\cr && +q \chi_{2z} (-103+360 \chi_{1x}^2-180 \chi_{1y}^2+q (5
   (-29+216 \chi_{1x}^2-120 \chi_{1y}^2)
\cr && -q (-27+252 q+670 q^2+348 q^3-1080 \chi_{1x}^2-360 q \chi_{1x}^2+60 (11+4 q) \chi_{1y}^2)\cr && -120 (1+q)^2 (1+3 q) \chi_{1x} \chi_{2x}-80 q (1+q)^2 (1+4 q) \chi_{2x}^2-120 (1+q)^2 \chi_{1y}
   \chi_{2y}\cr && +20 q (1+q)^2 (1+4 q) \chi_{2y}^2+20 q (1+q)^2 (1+4 q) \chi_{2z}^2))\bigg]\left(\frac{M}{r}\right)^{7/2}\Bigg\}.
\eea

The ADM mass of the system $M_{\rm ADM}$ expressed in term of the separation of the holes $r$ in ADMTT coordinates is given by
\bea
M \Omega &=&\frac{1}{384} \left(\frac{M}{r}\right)^{3/2} \Bigg\{384-\frac{192 (3+5 q+3 q^2)M}{(1+q)^2 r}-\frac{96 ((4+3 q) \chi_{1z}+q (3+4 q) \chi_{2z})}{(1+q)^2}\left(\frac{M}{r}\right)^{3/2}
\cr && -\frac{24}{(1+q)^4 } \bigg[16 (1+q)^2 (9+11 q+9 q^2)-2 (63+221 q+319 q^2+221 q^3+63 q^4)
\cr && -(1+q)^2 (6 (7-4 \chi_{1x}^2+2 \chi_{1y}^2+2 \chi_{1z}^2)
\cr && +q (41-48 \chi_{1x}
   \chi_{2x}+24 \chi_{1y} \chi_{2y}+24 \chi_{1z} \chi_{2z}+6 q (7-4 \chi_{2x}^2+2 \chi_{2y}^2+2 \chi_{2z}^2)))\bigg]\left(\frac{M}{r}\right)^{2}
\cr && +\frac{72 ((16+30
   q+34 q^2+13 q^3) \chi_{1z}+q (13+34 q+30 q^2+16 q^3) \chi_{2z})}{(1+q)^4}\left(\frac{M}{r}\right)^{5/2}
\cr && +\frac{1}{(1+q)^6} \bigg[48 (-10+38 \chi_{1x}^2-43 \chi_{1y}^2-7 \chi_{1z}^2)+4 q^4 (-10049+501 \pi ^2+930 \chi_{1x}^2-660 \chi_{1y}^2
\cr &&
-405
   \chi_{1z}^2 +5124 \chi_{1x} \chi_{2x}
+3546 \chi_{2x}^2-4872 \chi_{1y} \chi_{2y}-3216 \chi_{2y}^2-1914 \chi_{1z} \chi_{2z}
\cr && -1233 \chi_{2z}^2)+4 q^2 (-10049+501 \pi ^2+3546 \chi_{1x}^2-3216 \chi_{1y}^2-1233 \chi_{1z}^2+5124
   \chi_{1x} \chi_{2x}
\cr &&
+930 \chi_{2x}^2-4872 \chi_{1y} \chi_{2y}-660 \chi_{2y}^2-1914 \chi_{1z} \chi_{2z}-405 \chi_{2z}^2)
\cr && +48 q^6 (-10+38 \chi_{2x}^2-43 \chi_{2y}^2-7 \chi_{2z}^2)+q (501 \pi ^2+16 (-686+498
   \chi_{1x}^2-513 \chi_{1y}^2
\cr && -135 \chi_{1z}^2+360 \chi_{1x} \chi_{2x}-324 \chi_{1y} \chi_{2y}-144 \chi_{1z} \chi_{2z}))
\cr &&
+6 q^3 (501 \pi ^2+4 (-2470+490 \chi_{1x}^2-390 \chi_{1y}^2-197 \chi_{1z}^2
\cr && +1228 \chi_{1x}
   \chi_{2x}+490 \chi_{2x}^2-1192 \chi_{1y} \chi_{2y}-390 \chi_{2y}^2-446 \chi_{1z} \chi_{2z}-197 \chi_{2z}^2))
\cr && +q^5 (501 \pi ^2+16 (-686+360 \chi_{1x} \chi_{2x}
\cr && +498 \chi_{2x}^2-324 \chi_{1y} \chi_{2y}-513 \chi_{2y}^2-144
   \chi_{1z} \chi_{2z}-135 \chi_{2z}^2))\bigg]\left(\frac{M}{r}\right)^{3}
\cr && -\frac{12}{(1+q)^6} \bigg[12 (1+q)^2 (4+q) \chi_{1z}^3+36 q (1+q)^2 (3+2 q) \chi_{1z}^2 \chi_{2z}
\cr && +q \chi_{2z} (-27 (-5+8 \chi_{1x}^2-4
   \chi_{1y}^2)+q (385-648 \chi_{1x}^2+360 \chi_{1y}^2+72 \chi_{1x} \chi_{2x}+72 \chi_{1y} \chi_{2y})
\cr && +q^3 (377-216 \chi_{1x}^2+144 \chi_{1y}^2+504 \chi_{1x} \chi_{2x}+288 \chi_{2x}^2+72 \chi_{1y} \chi_{2y}
\cr && -72
   \chi_{2y}^2-72 \chi_{2z}^2)+9 q^4 (43+24 \chi_{1x} \chi_{2x}+48 \chi_{2x}^2-12 \chi_{2y}^2-12 \chi_{2z}^2)
\cr && +3 q^2 (121-216 \chi_{1x}^2+132 \chi_{1y}^2+120 \chi_{1x} \chi_{2x}
\cr && +16 \chi_{2x}^2+48 \chi_{1y} \chi_{2y}-4
   \chi_{2y}^2-4 \chi_{2z}^2)+24 q^5 (7+8 \chi_{2x}^2-2 \chi_{2y}^2-2 \chi_{2z}^2))
\cr && +\chi_{1z} (24 (7-8 \chi_{1x}^2+2 \chi_{1y}^2)-9 q (-43+48 \chi_{1x}^2-12 \chi_{1y}^2+24 \chi_{1x}
   \chi_{2x})
\cr && +27 q^5 (5+8 \chi_{2x}^2-4 \chi_{2y}^2-4 \chi_{2z}^2)-q^2 (-377+288 \chi_{1x}^2-72 \chi_{1y}^2+504 \chi_{1x} \chi_{2x}
\cr && -216 \chi_{2x}^2+72 \chi_{1y} \chi_{2y}+144 \chi_{2y}^2+72 \chi_{2z}^2)-3 q^3
   (-121+16 \chi_{1x}^2-4 \chi_{1y}^2
\cr && +120 \chi_{1x} \chi_{2x}-216 \chi_{2x}^2+48 \chi_{1y} \chi_{2y}+132 \chi_{2y}^2+84 \chi_{2z}^2)
\cr && -q^4 (-385+72 \chi_{1x} \chi_{2x}-648 \chi_{2x}^2+72 \chi_{1y} \chi_{2y}+360 \chi_{2y}^2+288
   \chi_{2z}^2))\bigg]\left(\frac{M}{r}\right)^{7/2}\Bigg\}.
\eea

The orbital angular frequency $M\Omega$ expressed in term of the separation of the holes $r$ in ADMTT coordinates is given by
\bea
\frac{M_{\rm ADM}}{M} &=&1 +  \frac{M}{384 r} \Bigg\{-192+\frac{48 (7+q (13+7 q)) M}{(1+q)^2 r}
-\frac{96 ((4+3 q) \chi_{1z}+q (3+4 q) \chi_{2z})}{(1+q)^2} \left(\frac{M}{r}\right)^{3/2}
\cr && +\frac{24}{(1+q)^4 } \bigg[9-8 \chi_{1x}^2+4 \chi_{1y}^2+4 \chi_{1z}^2+q^4 (9-8 \chi_{2x}^2+4 \chi_{2y}^2+4 \chi_{2z}^2)
\cr && +8 q (2-2 \chi_{1x} (\chi_{1x}+\chi_{2x})+\chi_{1y} (\chi_{1y}+\chi_{2y})\cr && +\chi_{1z}
   (\chi_{1z}+\chi_{2z}))+8 q^3 (2-2 \chi_{2x} (\chi_{1x}+\chi_{2x})+\chi_{2y} (\chi_{1y}+\chi_{2y})+\chi_{2z} (\chi_{1z}+\chi_{2z}))
\cr && +q^2 (13 -8 (\chi_{1x}^2+4 \chi_{1x} \chi_{2x}+\chi_{2x}^2)+4 (\chi_{1y}^2+4 \chi_{1y}
   \chi_{2y}+\chi_{2y}^2)
\cr && +4 (\chi_{1z}^2+4 \chi_{1z} \chi_{2z}+\chi_{2z}^2))\bigg]\left(\frac{M}{r}\right)^{2}
\cr && -\frac{24 ((32+q (42+q (14+q))) \chi_{1z}+q (1+2 q (7+q (21+16 q)))
   \chi_{2z})}{(1+q)^4}\left(\frac{M}{r}\right)^{5/2}
\cr && +\frac{1}{(1+q)^6} \bigg[537-1248 \chi_{1x}^2+48 \chi_{1y}^2+1200 \chi_{1z}^2+q (-3497+243 \pi ^2 (1+q)^4
\cr && -2784 \chi_{1x}^2-144
   \chi_{1y} (5 \chi_{1y}+8 \chi_{2y})+144 \chi_{1z} (23 \chi_{1z}+10 \chi_{2z})
\cr && +3 q^2 (-9787+304 \chi_{1x}^2+352 \chi_{1z}^2+288 \chi_{1x} \chi_{2x}+304 \chi_{2x}^2
\cr && -16 (47 \chi_{1y}^2+156 \chi_{1y} \chi_{2y}+47
   \chi_{2y}^2)+2976 \chi_{1z} \chi_{2z}+352 \chi_{2z}^2)
\cr && +3 q^5 (179-416 \chi_{2x}^2+16 \chi_{2y}^2+400 \chi_{2z}^2)-q^4 (3497+2784 \chi_{2x}^2+144 \chi_{2y} (8 \chi_{1y}+5 \chi_{2y})\cr && -144 \chi_{2z} (10 \chi_{1z}+23
   \chi_{2z}))+q (-18707-1224 \chi_{1x}^2+432 \chi_{1x} \chi_{2x}+600 \chi_{2x}^2
\cr && -144 (16 \chi_{1y}^2+34 \chi_{1y} \chi_{2y}+5 \chi_{2y}^2)+72 (43 \chi_{1z}^2+82 \chi_{1z} \chi_{2z}+\chi_{2z}^2))
\cr && +q^3
   (-18707+600 \chi_{1x}^2+432 \chi_{1x} \chi_{2x}-1224 \chi_{2x}^2-144 (5 \chi_{1y}^2+34 \chi_{1y} \chi_{2y}+16 \chi_{2y}^2)
\cr && +72 (\chi_{1z}^2+82 \chi_{1z} \chi_{2z}+43 \chi_{2z}^2)))\bigg]\left(\frac{M}{r}\right)^{3}
\cr && -\frac{6}{(1+q)^6} \bigg[12 (1+q)^2 (12+5 q) \chi_{1z}^3+12 q (1+q)^2 (29+22 q) \chi_{1z}^2 \chi_{2z}
\cr && +q \chi_{2z} (128-504 \chi_{1x}^2+252 \chi_{1y}^2+q (181-1536 \chi_{1x}^2+24 \chi_{1y} (35 \chi_{1y}+9
   \chi_{2y})
\cr && +4 q^3 (136+30 \chi_{2x} (2 \chi_{1x}+5 \chi_{2x})+3 (8 \chi_{1y}-7 \chi_{2y}) \chi_{2y}-21 \chi_{2z}^2)
\cr && +3 q^2 (27-176 \chi_{1x}^2+112 \chi_{1y}^2+160 \chi_{1x} \chi_{2x}+112 \chi_{2x}^2+136 \chi_{1y} \chi_{2y}-8
   \chi_{2y}^2-8 \chi_{2z}^2)
\cr && +24 q^4 (13+12 \chi_{2x}^2-2 \chi_{2y}^2-2 \chi_{2z}^2)
+4 q (-22+6 (-65 \chi_{1x}^2+10 \chi_{1x} \chi_{2x}+\chi_{2x}^2)
\cr && +3 (77 \chi_{1y}^2+44 \chi_{1y}
   \chi_{2y}+\chi_{2y}^2)+3 \chi_{2z}^2)))+\chi_{1z} (24 (13-20 \chi_{1x}^2+6 \chi_{1y}^2)
\cr && +q (544+348 \chi_{1y}^2-24 \chi_{1x} (47 \chi_{1x}+26 \chi_{2x})+96 \chi_{1y} \chi_{2y}+4 q^4 (32+90
   \chi_{2x}^2-45 \chi_{2y}^2-21 \chi_{2z}^2)
\cr && -3 q (-27+272 \chi_{1x}^2-88 \chi_{1y}^2+512 \chi_{1x} \chi_{2x}-112 \chi_{2x}^2-40 \chi_{1y} \chi_{2y}+80 \chi_{2y}^2+8 \chi_{2z}^2)
\cr && -4 q^2 (22+42 \chi_{1x}^2-15
   \chi_{1y}^2+300 \chi_{1x} \chi_{2x}-258 \chi_{2x}^2 +12 \chi_{1y} \chi_{2y}+165 \chi_{2y}^2+33 \chi_{2z}^2)
\cr && -q^3 (-181+96 (3 \chi_{1x}-11 \chi_{2x}) \chi_{2x}+72 \chi_{1y} \chi_{2y}+600 \chi_{2y}^2+192
   \chi_{2z}^2)))\bigg]\left(\frac{M}{r}\right)^{7/2}\Bigg\}.
\eea
\end{widetext}

\bibliographystyle{apsrev4-1}
\bibliography{../../../../Bibtex/references.bib}

\end{document}